# Climate benefits of afforestation and reforestation with varying species mixtures and densities in the north-western boreal lands

Enoch Ofosu[a*], Kevin Bradley Dsouza[a], Daniel Chukwuemeka Amaogu[d], Jérôme Pigeon[d], Richard Boudreault[c, a], Juan Moreno-Cruz[b*], Pooneh Maghoul[e,d], Yuri Leonenko[a*]

* Corresponding authors
a -
b - School of Environment, Enterprise and Development, University of Waterloo
c- AWN Nanotech
d- Department of Civil, Geological and Mining Engineering, Polytechnique Montréal
e- The United Nations University Institute for Water, Environment and Health (UNU-INWEH)

## Abstract

The boreal forest is a vital carbon sink. Using 250‑year simulations for Canada's Taiga Plains, a priority of the 2 Billion Trees Program, we tested afforestation/reforestation strategies that combine species mix, planting density and surface albedo. Medium‑density (600–1400 trees $ha^{-1}$) mixed stands with ~25–40 % deciduous trees stored 15–30 % more net ecosystem carbon than conifer monocultures by coupling rapid early growth with long‑term retention and greater disturbance resilience. Replanting under‑stocked stands with these mixtures raised long‑term storage by 18–30 % over business‑as‑usual. Accounting for albedo showed pure evergreen or deciduous stands lost 6–20 % of their climate benefit, whereas mixed stands yielded net cooling and the highest sequestration (≈ 4.6–4.7 t $CO_2$e $ha^{-1}$ $yr^{-1}$). Partial harvesting plus replanting preserved, and sometimes increased, ecosystem carbon (≈ 300–340 t C $ha^{-1}$) and productivity (≈ 1.6–2.0 t C $ha^{-1}$ $yr^{-1}$) without raising risk. Blending fast‑growing deciduous trees with long‑lived conifers at intermediate density maximizes boreal A/R climate value and informs reforestation policy elsewhere.

## Introduction

Afforestation and reforestation (A/R) have gained recognition as cost-effective Natural Climate Solutions (NCS) to mitigate global climate change, including in high-latitude ecosystems [1, 2]. In Canada, the 2 Billion Trees Program is a cornerstone of the federal government's net-zero strategy [3], emphasizing the need to strategically reforest marginal lands with species and configurations that optimize carbon sequestration [4, 5]. The Taiga Plains ecoregion, spanning over 570,000 km [2], is uniquely positioned to benefit from well-planned A/R due to its extensive permafrost landscapes, substantial peatland carbon reserves, and rapid climatic changes [6, 7, 16]. Warming trends exceeding 2 – 4 °C since 1950 [19] have contributed to thawing permafrost, altered snow regimes, intensified fire cycles, and the northward migration of deciduous tree species such as aspen [9, 10, 13]. These transitions underscore the importance of adaptive reforestation strategies that integrate climate trends, species dynamics, and landscape heterogeneity [11, 12].

Previous studies have demonstrated that coniferous species such as black spruce (*Picea mariana*) exhibit strong carbon retention and stress tolerance but grow slowly [14, 15], while deciduous species like aspen (*Populus tremuloides*) promote rapid early carbon uptake but have shorter life spans and greater disturbance susceptibility [13,14]. Yet, the optimal configuration of mixed-species assemblages and planting densities that balance these trade-offs in boreal landscapes remains poorly resolved [8,18]. Moreover, boreal afforestation can generate biophysical climate feedbacks such as reduced albedo, which may offset carbon sequestration gains, particularly during snow-covered months [20, 26, 30]. These complex dynamics, including key uncertainties, are illustrated in Fig. 1 (b), (c). These albedo trade-offs are rarely included in large-scale forest carbon assessments despite their growing policy relevance [27, 29, 37]. To address these gaps (summarized visually in Fig. 1(c)), we employ an integrated modeling approach using the Growth and Yield Projection System (GYPSY) [23] and the Carbon Budget Model of the Canadian Forest Sector (CBM-CFS3) [24] to evaluate 250-year carbon stock trajectories under various A/R scenarios in the Taiga Plains. Surface albedo changes in each scenario were translated into radiative forcing using monthly albedo outputs and radiative kernels [25], and further expressed as an equivalent $CO_2$ flux offset using established methods [26]. Our analysis focuses on net carbon sequestration and albedo-driven radiative effects; other important processes – such as atmospheric energy redistribution and permafrost-carbon feedback – are beyond the scope of this study.

Our main objective in this study is to assess the net climate benefit of boreal afforestation and reforestation across species mixtures and planting densities. Specifically, we accomplish the following: a) Quantify the net ecosystem carbon (NEC) of monoculture and mixed-species stands under low, medium, and high planting densities. b) Determine the carbon sequestration gains from reforesting currently under-stocked (≤ 600 trees/ha) forests using various species combinations. c) Evaluate the effects of strategic harvesting with subsequent replanting on long-term carbon dynamics. d) Incorporate surface albedo effects into net climate benefit calculations. By integrating biogeochemical and biophysical climate drivers, this study provides actionable insights for Canada's 2 Billion Trees Program and other boreal forest-based climate initiatives, ensuring that A/R strategies maximize carbon sequestration while avoiding unintended positive climate feedback.

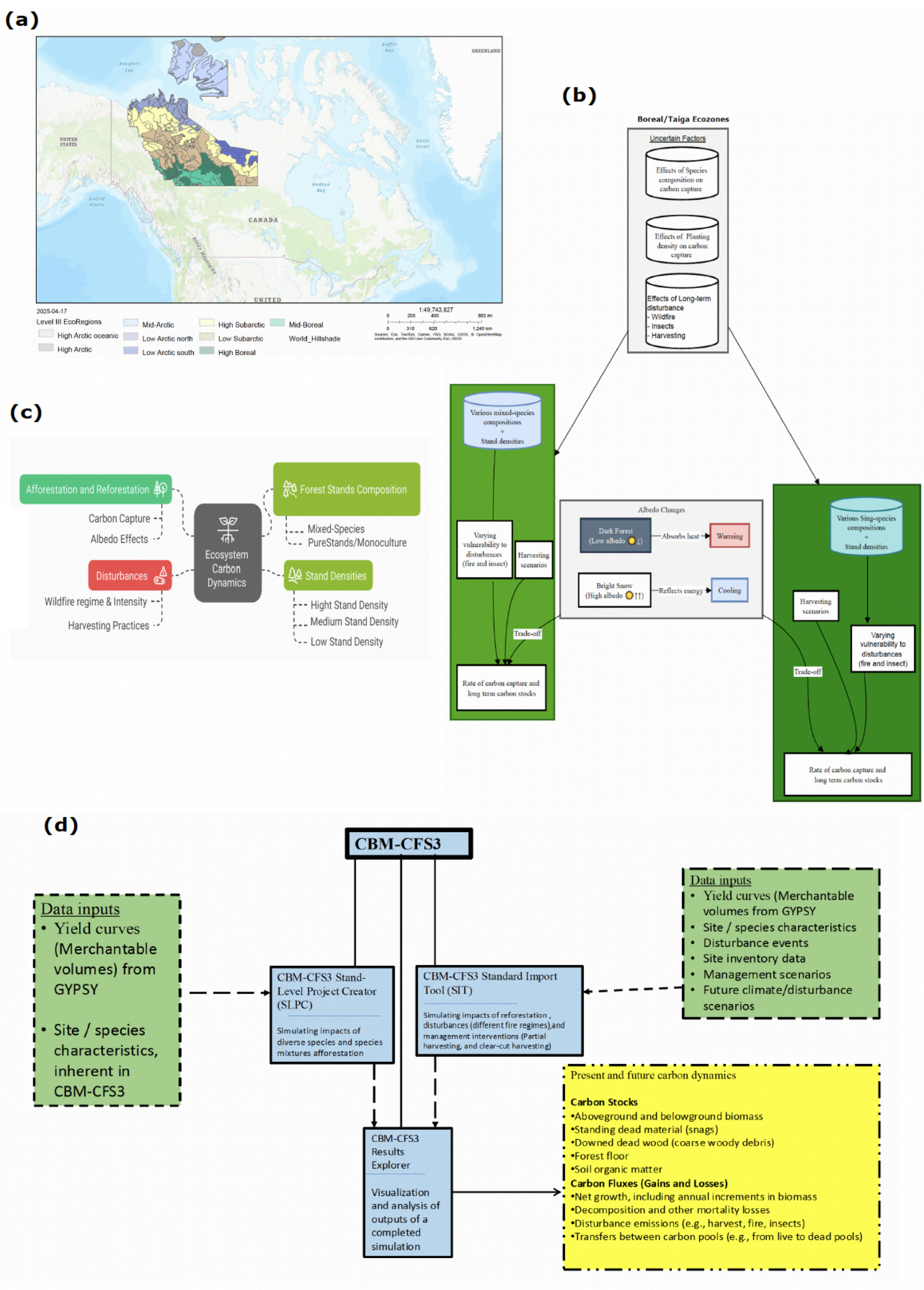

(a)
GREENLAND
CANADA
Rocky Mountains
UNITED
2025-04-17
Level III EcoRegions
High Arctic oceanic
High Arctic
Mid-Arctic
Low Arctic north
Low Arctic south
High Subarctic
Low Subarctic
High Boreal
Mid-Boreal
World_Hillshade
1:49,743,827
0 200 400 800 mi
0 310 620 1,240 km
(b)
Boreal/Taiga Ecozones
Uncertain Factors
Effects of Species composition on carbon capture
Effects of Planting density on carbon capture
Effects of Long-term disturbance
- Wildfire
- Insects
- Harvesting
Various mixed-species compositions + Stand densities
Varying vulnerability to disturbances (fire and insect)
Harvesting scenarios
Trade-off
Rate of carbon capture and long term carbon stocks
Albedo Changes
Dark Forest (Low albedo ↓)
Absorbs heat
Warming
Bright Snow (High albedo ↑↑)
Reflects energy
Cooling
Various Sing-species compositions + Stand densities
Harvesting scenarios
Varying vulnerability to disturbances (fire and insect)
Trade-off
Rate of carbon capture and long term carbon stocks
(c)
Afforestation and Reforestation
Carbon Capture
Albedo Effects
Disturbances
Wildfire regime & Intensity
Harvesting Practices
Ecosystem Carbon Dynamics
Forest Stands Composition
Mixed-Species
PureStands/Monoculture
Stand Densities
Hight Stand Density
Medium Stand Density
Low Stand Density
(d)
CBM-CFS3
Data inputs
• Yield curves (Merchantable volumes) from GYPSY
• Site / species characteristics, inherent in CBM-CFS3
CBM-CFS3 Stand-Level Project Creator (SLPC)
Simulating impacts of diverse species and species mixtures afforestation
CBM-CFS3 Standard Import Tool (SIT)
Simulating impacts of reforestation , disturbances (different fire regimes),and management interventions (Partial harvesting, and clear-cut harvesting)
Data inputs
• Yield curves (Merchantable volumes from GYPSY
• Site / species characteristics
• Disturbance events
• Site inventory data
• Management scenarios
• Future climate/disturbance scenarios
CBM-CFS3 Results Explorer
Visualization and analysis of outputs of a completed simulation
Present and future carbon dynamics
Carbon Stocks
•Aboveground and belowground biomass
•Standing dead material (snags)
•Downed dead wood (coarse woody debris)
•Forest floor
•Soil organic matter
Carbon Fluxes (Gains and Losses)
•Net growth, including annual increments in biomass
•Decomposition and other mortality losses
•Disturbance emissions (e.g., harvest, fire, insects)
•Transfers between carbon pools (e.g., from live to dead pools)

**Fig. 1: Panels Overview: Integrating Boreal Afforestation Dynamics and Carbon Modeling Frameworks.** (a) Geographic Scope of Study in Northern Canada. This panel shows the location of the study area across key ecozones in northern Canada, including High Boreal, Mid-Boreal, and Subarctic zones, forming the spatial domain for afforestation and reforestation simulations. (b) Key Drivers of Ecosystem Carbon Dynamics. This schematic presents how afforestation/reforestation practices, disturbance regimes, stand composition, and planting density interact to influence carbon capture and albedo responses within boreal ecosystems.(c) Conceptual Uncertainties in Afforestation Outcomes. This diagram highlights uncertainties related to species mixtures, disturbance responses, and long-term ecosystem feedbacks, complicating projections of afforestation benefits in boreal and taiga zones.(d) Structure of the CBM-CFS3 Carbon Accounting Model. This flowchart outlines the internal logic of the Carbon Budget Model of the Canadian Forest Sector (CBM-CFS3), including input data types, Growth and yield curves generated with the Growth and Yield Projection System (GYPSY), disturbance modules, and analytical outputs used for evaluating forest carbon dynamics over time.

# Results

Our simulations for the Taiga Plains ecoregion reveal distinct carbon sequestration patterns across different species compositions and planting densities [low (≤600 trees/ha), medium (600–1400 trees/ha), and high (1400–2500 trees/ha)] over a 250-year period (Figure 2).

### Influence of Species Mix and Density on Carbon Sequestration Trajectories

Mixed-species stands, particularly the "AllMix" scenario (25% each of black spruce, white spruce, pine, and aspen), generally demonstrated superior long-term Net Ecosystem Carbon (NEC) accumulation compared to most monoculture-dominated scenarios across all densities (Figure 2i). At the 250-year mark, the AllMix scenario achieved the highest NEC values, reaching approximately 230 tC $ha^{-1}$ at low density, 270 $tCha^{-1}$ at medium density, and 260 tC $ha^{-1}$ at high density (Fig. 2. ii (a, b, c)). This enhanced performance appears driven by ecological complementarity, where the rapid early growth of deciduous species like aspen is effectively balanced by the long-term carbon retention of conifers such as black spruce [13, 15] (Fig. 2. i). Notably, these mixed stands are also suggested to exhibit greater resilience to disturbances, a critical factor for maintaining stable carbon stocks under changing climatic conditions [10, 22]. For instance, at medium density, the NEC of AllMix at year 250 was significantly greater than that of 95%AW (approx. 190 t C $ha^{-1}$), 95%SB (approx. 180 tC $ha^{-1}$), and 95%SW (approx. 140 $tCha^{-1}$) (Fig. 2. ii (b)).

Monoculture-dominated plantings displayed varied carbon accumulation trajectories (Fig. 2. i). Aspen-dominated stands (e.g., 95%AW, 40%AW) exhibited the most rapid carbon gains in the initial decades. For example, across all densities, the 95%AW stands achieved significantly higher NEC by year 100 (e.g., ~160 tC $ha^{-1}$ at medium density) compared to spruce-dominated stands like 95%SB (e.g., ~130 t C $ha^{-1}$ at medium density) (Figure 2i). However, this early advantage of aspen was transient, with its NEC accumulation rates declining and plateauing after approximately 100-150 years (Figure 2i), partly due to its shorter lifespan and higher susceptibility to disturbances [14]. By year 250, aspen-dominated stands like 95%AW showed NEC values of around 180-200 tC $ha^{-1}$ depending on density (Fig. 2. ii). In contrast, spruce-dominated stands (e.g., 95%SB, 95%SW) showed slower initial accumulation but more sustained carbon uptake into later centuries (Fig. 2.i). Black spruce, for instance, demonstrated remarkable stability in later centuries, consistent with its stress-tolerant growth strategy [13, 15], although its overall NEC at year 250 was generally lower than AllMix and aspen-dominated stands (e.g., at medium density, 95%SB ~180 t C $ha^{-1}$, 95%SW ~140 tC $ha^{-1}$) (Figure 2ii). White spruce (SW) generally showed intermediate performance between aspen and black spruce in the long term (Fig. 2. i, 2.ii).

Planting density was a critical factor influencing both the rate and total amount of carbon accumulation (Figure 2i, 2ii). High-density stands (1400–2500 trees/ha) consistently achieved the highest NEC values in the first 50-100 years across most species compositions. For example, the AllMix scenario at high density reached approximately 130 tC $ha^{-1}$ by year 50, compared to ~100 tC $ha^{-1}$ for AllMix at medium density (Figure 2i b, c). However, this initial advantage of high density often diminished beyond 100 years, potentially due to competition-induced mortality and accelerated organic matter decomposition [23, 28], and was surpassed by medium-density stands by year 250 for several mixtures. Medium-density stands (600–1400 trees/ha) proved optimal for long-term NEC, particularly for the AllMix scenario, which achieved its peak NEC of approximately 270 tC $ha^{-1}$ at this density (Fig. 2.ii (b)). This was slightly higher than AllMix at high density (~260 t C $ha^{-1}$, Fig. 2.ii (c) ) and significantly higher than many high-density monocultures (e.g., the NEC for AllMix at medium density was significantly greater than 40%SW at high density, as shown in Fig. 2. iii (b, c)). Low-density stands (≤ 600 trees/ha) consistently resulted in the lowest NEC accumulation across all species configurations and throughout the 250-year simulation (Fig. 2.i (a), 2.ii (a)).

These results from Fig. 2. collectively indicate that medium-density afforestation strategies incorporating a balanced mix of deciduous and coniferous species, such as the AllMix scenario, provide the most robust approach for maximizing long-term NEC in the Taiga Plains. This combination appears to effectively balance rapid early carbon uptake with sustained long-term storage, outperforming both high-density plantings and most monoculture scenarios by the end of the 250-year simulation period. These findings offer valuable insights for guiding afforestation programs to optimize both climate mitigation and forest productivity objectives [3, 18].

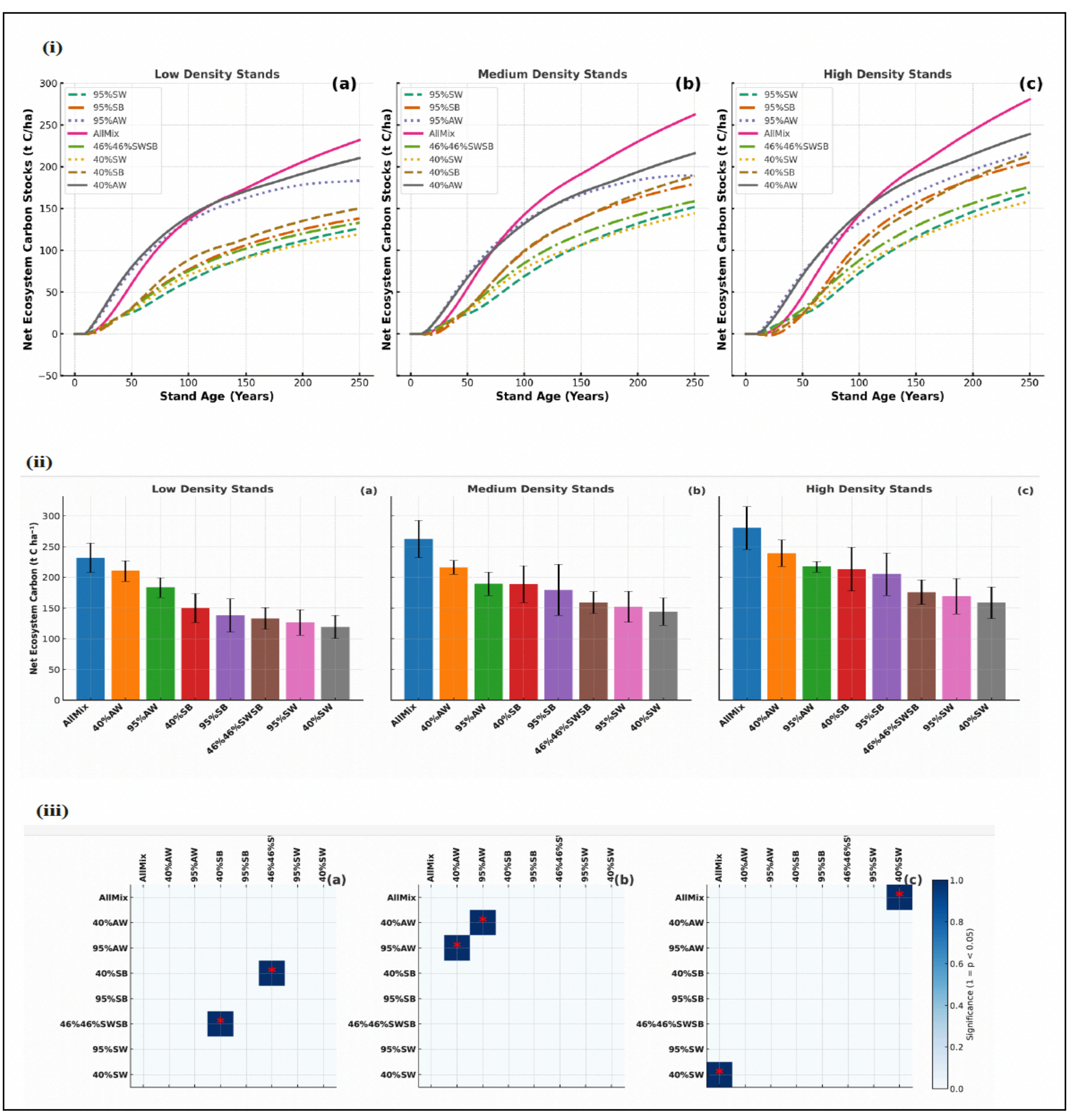

**Fig. 2. Carbon capture projections and statistical comparisons for afforestation scenarios in the Taiga Plains region.** (i) Projected Net Ecosystem Carbon Stocks (tC $ha^{-1}$) over 250 years for afforestation with various tree species and combinations under (a) low (≤ 600 trees/ha), (b) medium (600 – 1400 trees/ha), and (c) high (1400 – 2500 trees/ha) stand densities. Abbreviations: SW = White spruce (*Picea glauca*), SB = Black spruce (*Picea mariana*), AW = Aspen (*Populus tremuloides*), AllMix = balanced mixed species stand (25% SB, 25% SW, 25% Pine, 25% AW). (ii) Net Ecosystem Carbon Stocks (NEC, $tCha^{-1}$) at Year 250 under varying species mixtures and stand densities: (a) low, (b) medium, and (c) high. Bars represent NEC for different monoculture and mixed-species scenarios, including AllMix (MX: 25% SB, 25% SW, 25% Pine, 25% AW), 40%AW, 95%AW, 40%SB, 95%SB, 46%46%SWSB (a mix of White Spruce and Black Spruce), 95%SW, and 40%SW. Error bars indicate standard deviation. (iii) Matrix visualization of statistical significance (pairwise t-test, $p < 0.05$, with Tukey HSD post-hoc correction) for Net Ecosystem Carbon among all species compositions at Year 250 under (a) low, (b) medium, and (c) high tree density stands. Species compositions are listed along both axes. A blue cell with a bold red asterisk (*) indicates a statistically significant difference ($p < 0.05$) between the corresponding species compositions. The color bar indicates the significance level (1 - p-value).

**Reforestation on Low-Density Forests**

The projected Net Ecosystem Carbon Stocks (NEC, in tC ha$^{-1}$) over a 250-year period for four primary Taiga Plains site types [(a) Mixedwood, (b) Aspen, (c) White Spruce, and (d) Black Spruce], illustrated in Fig.3. (i). These simulations compare a Business-As-Usual (BAU) scenario, which represents an initial low tree stand density ($\leq$ 600 trees/ha), with various reforestation treatments designed to restore these sites to a higher tree density (1400 – 2500 trees/ha) using different species. The reforestation treatments included Aspen (AW), White Spruce (SW), and Black Spruce (SB). A mixed-species treatment (MX: 25% SB, 25% SW, 25% Pine, 25% AW) was exclusively evaluated on the Mixedwood site under this high-density planting regime. Fig. 3. (ii) presents the corresponding pairwise statistical significance matrices, indicating where NEC at year 250 differed significantly ($p < 0.05$, denoted by red asterisks) between treatments for each site type.

On the Mixedwood Site (Fig. i-a), the MX treatment, when applied to restore the low-density stand, was projected to achieve the highest NEC, accumulating approximately 250 tC ha$^{-1}$ by year 250. This performance aligns with principles of ecological complementarity [13, 15], where species mixtures can enhance resource utilization. The AW treatment followed with ~230 tC ha$^{-1}$, SW with ~190 tCha$^{-1}$, and SB with ~170 tC ha$^{-1}$, all under the high-density reforestation. The BAU scenario, representing the un-restored low-density stand, resulted in the lowest NEC, around 50 tC ha$^{-1}$. The pairwise statistical significance matrix (Fig. ii-a) confirms that at year 250, NEC under the MX treatment was significantly higher ($p < 0.05$) than under BAU, SB, and SW treatments. However, the difference between MX and AW was not statistically significant. The AW treatment, in turn, showed significantly higher NEC compared to BAU, SB, and SW ($p < 0.05$). No significant difference was observed between the NEC of SB and SW treatments, though both were significantly higher than BAU ($p < 0.05$).

For the Aspen Site (Fig. i-b), aspen treatment (AW) used for high-density reforestation demonstrated the most rapid initial carbon accumulation, outperforming other treatments for approximately the first 100 years. By year 250, however, the SW monoculture projected the highest NEC (~200 tC ha$^{-1}$), followed closely by SB (~180 tCha$^{-1}$) and then AW (~170 tC ha$^{-1}$). The BAU scenario (low-density) showed markedly lower NEC ($<$25 tCha$^{-1}$). Despite these numerical differences in final NEC among the high-density reforestation treatments, the pairwise statistical significance matrix (Fig. ii-b) indicated no significant differences ($p > 0.05$) among the AW, SW, and SB treatments at year 250. All three reforestation treatments, however, resulted in significantly higher NEC than the BAU scenario ($p < 0.05$).

On the White Spruce Site (Fig. i-c), SW treatment, when used for high-density reforestation, were projected to achieve the highest NEC by year 250, accumulating approximately 220 tCha$^{-1}$. The AW treatment followed with ~190 tCha$^{-1}$, and the SB treatment with ~140 tC ha$^{-1}$. The BAU (low-density) scenario resulted in an NEC of around 40 tC ha$^{-1}$. The pairwise statistical significance matrix for the White Spruce site (Fig. ii-d) revealed that all pairwise comparisons of NEC among AW, SW, SB reforestation treatments and the BAU scenario were statistically significant ($p < 0.05$), with the ranking of NEC being SW > AW > SB > BAU.

For the Black Spruce Site (Fig. i-d), SB treatment applied in the high-density reforestation exhibited the highest projected NEC at year 250, approximately 200 tC ha$^{-1}$, underscoring SB's niche adaptation to such environments [13, 26]. The SW treatment accumulated ~170 tCha$^{-1}$, and the AW treatment ~130 tC ha$^{-1}$. The BAU (low-density) scenario showed the lowest NEC, around 30 tC ha$^{-1}$.

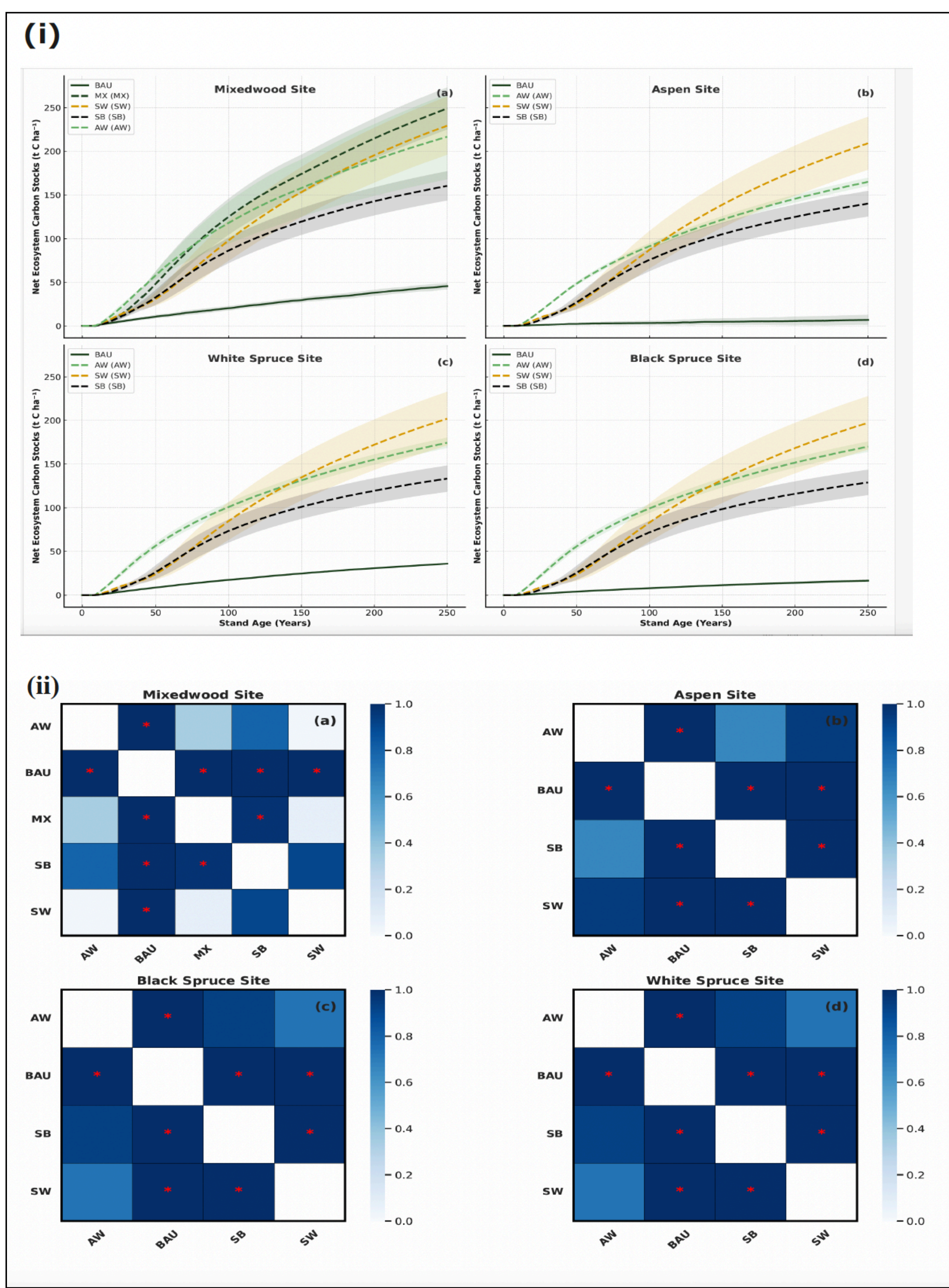

**Fig. 3. Projected carbon capture and statistical comparisons of reforestation strategies across four boreal forest site types in the Taiga Plains.** (i) Projected net ecosystem carbon stocks (tC $ha^{-1}$) under various reforestation scenarios on low-density sites in the Taiga Plains, evaluated across four site types: (a) Mixedwood site, (b) Aspen site, (c) White spruce site, and (d) Black spruce site. Simulations are based on Business-as-Usual (BAU) disturbance regimes and include different planting strategies: Mixed-species (MX: 25% Black spruce (SB, *Picea mariana*), 25% White spruce (SW, *Picea glauca*), 25% Pine, 25% Aspen (AW, *Populus tremuloides*)), White spruce (SW), Black spruce (SB), and Aspen (AW). Stand densities modeled include low ($\leq 600$ trees/ha), medium (600 – 1400 trees/ha), and high (1400 – 2500 trees/ha). The shaded regions around each line represent standard deviation, indicating variability in projected carbon accumulation across replicates. (ii) Pairwise statistical significance matrix (1 - p-value) of Net Ecosystem Carbon Stocks (tC $ha^{-1}$) at Year 250 across all reforestation species and Business-As-Usual (BAU) scenarios for each site type: (a) Mixedwood, (b) Aspen, (c) Black spruce, and (d) White spruce. Each matrix visualizes the pairwise treatment comparisons using Tukey HSD post-hoc test after ANOVA. Blue shading indicates higher statistical significance (i.e., lower p-values, higher 1 - p-value). A bold red asterisk (*) marks pairs with significant differences ($p < 0.05$).

Similar to the White Spruce site, the statistical matrix for the Black Spruce site (Fig. ii-c) confirmed that all pairwise differences in NEC among the AW, SW, SB reforestation treatments and the BAU scenario were statistically significant ($p < 0.05$), with the NEC order being SB > SW > AW > BAU.

Across all site types and active reforestation treatments (MX, AW, SW, SB), variability in projected NEC outcomes is represented by the shaded bands in Figure (i). It is important to note that the mixed-species (MX) treatment was only simulated for the Mixedwood site, thus precluding direct graphical or statistical comparison of its performance on the Aspen, White Spruce, or Black Spruce sites when restoring low-density stands to high density.

**Pathways to Carbon Resilience: Integrating Harvesting and Reforestation Post-Disturbance**

Fig.4. illustrates that strategies involving partial harvesting followed by planting (e.g., BAU_PH_Planting) are projected to maintain higher ecosystem carbon stocks by year 250 (approximately 275 tC $ha^{-1}$) compared to clearcut and replanting strategies (e.g., BAU_ClearCut_Planting, approximately 250 tC $ha^{-1}$). This is partly because partial harvesting results in a less severe initial decline in Net Ecosystem Productivity (NEP) immediately post-harvest (Fig.4(a)), where BAU_PH_Planting shows a shallower dip than BAU_ClearCut_Planting. Its benefits, such as maintaining biodiversity and enabling sustainable timber production, extend beyond carbon metrics alone [34, 47].

Conversely, while clearcut-replanted stands (e.g., BAU_ClearCut_Planting) exhibit a more pronounced initial drop in both NEP (Fig.4(a)), dipping to nearly -0.25 tC $ha^{-1}$ $yr^{-1}$ within the first 50 years) and total ecosystem carbon stocks (Fig.4(b)), falling below 225 tC $ha^{-1}$ initially), the simulations indicate that with robust replanting, these stands show a strong recovery. NEP in the BAU_ClearCut_Planting scenario rebounds significantly, peaking above 1.0 tC $ha^{-1}$ $yr^{-1}$ around year 100-125 (Fig.4(a)), leading to substantial carbon stock recovery by year 250. Thus, they remain a viable strategy for rapid wood supply without catastrophic long-term carbon losses relative to scenarios involving future fire impacts (e.g., FFS_ClearCut_Planting, which results in stocks around 220 tC $ha^{-1}$ by year 250), provided robust replanting protocols are followed [5, 9].

Policymakers and managers should interpret these trends as evidence that multiple pathways, ranging from partial (BAU_PH_Planting) to clearcut harvesting (BAU_ClearCut_Planting), when paired with deliberate replanting, can achieve significant carbon outcomes, though partial harvesting demonstrates a modest carbon advantage in these long-term simulations. By year 250, both BAU_PH_Planting and BAU_ClearCut_Planting lead to ecosystem carbon stocks that are considerably higher than those projected for stands affected by future fire scenarios with subsequent planting (FFS_PH_Planting and FFS_ClearCut_Planting, both ending around 220-230 tC $ha^{-1}$) and even surpass the unharvested Future Fire Scenario (FFS, ending ~240 tC $ha^{-1}$) (Fig.4. (b)). The choice between these harvesting approaches, when coupled with replanting, may therefore depend on broader ecological, economic, and social priorities [3, 22]. This insight aligns with the principles of diversified risk management, critical in an era of escalating climatic uncertainty [10, 20].

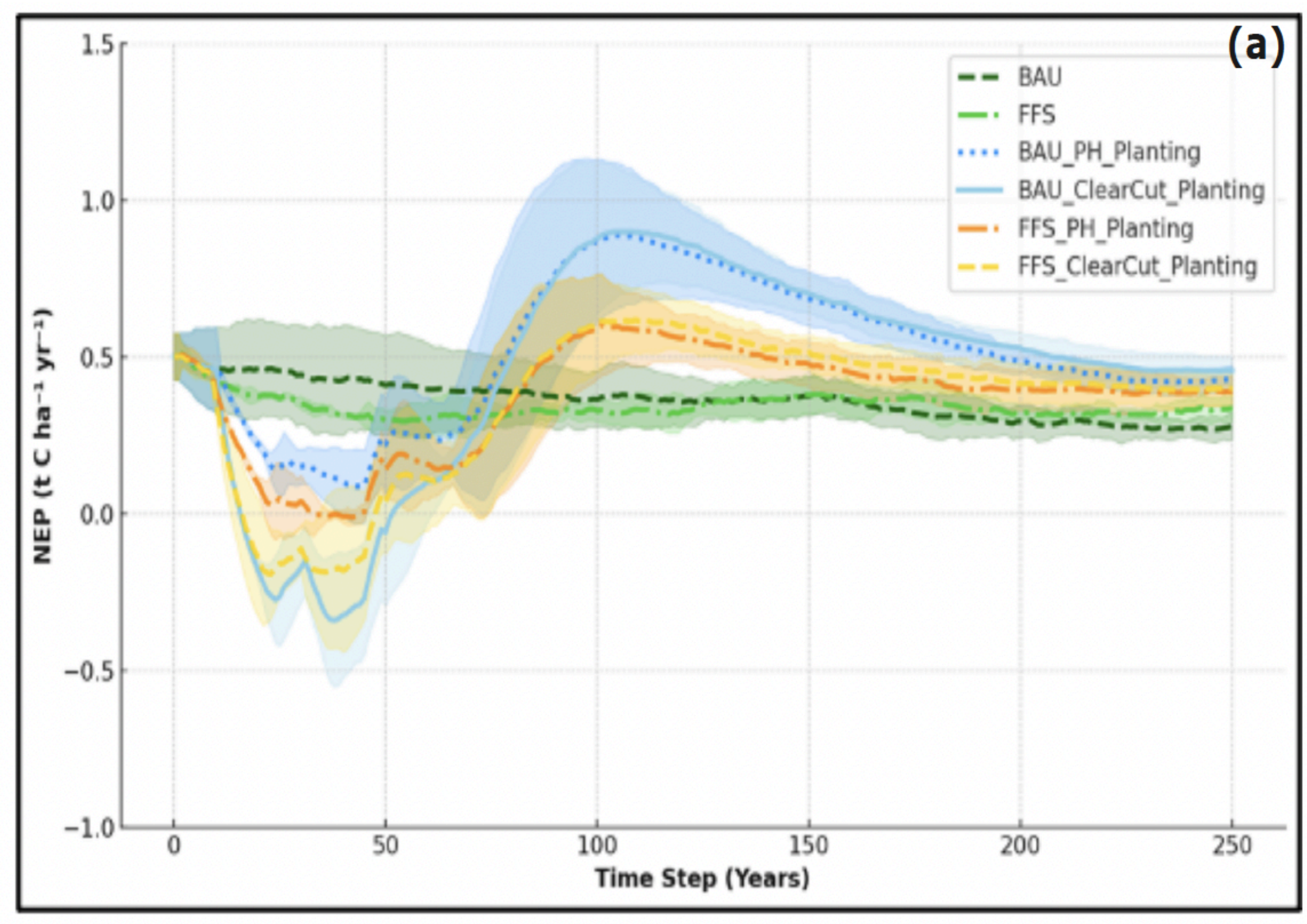

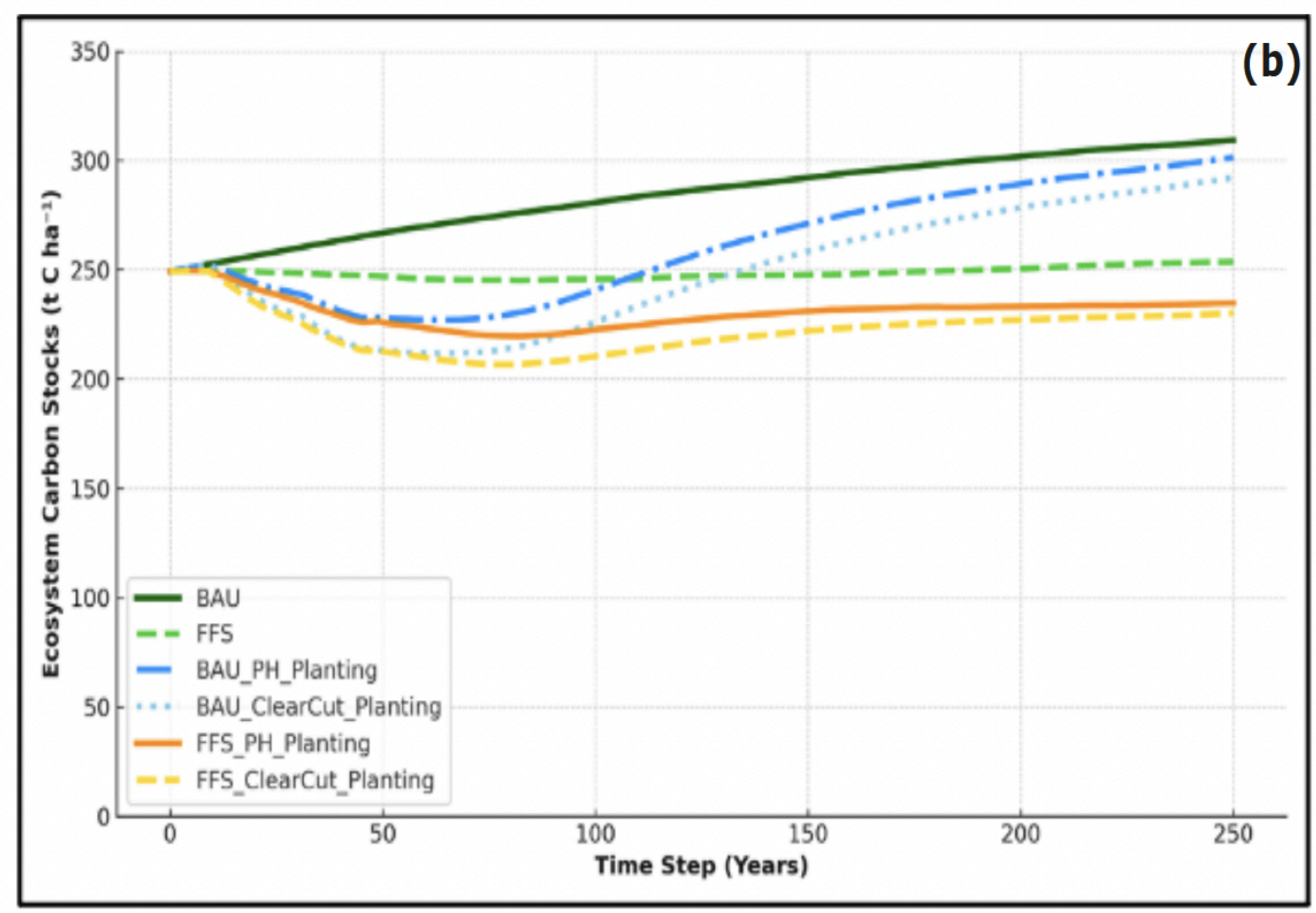

**Fig.4. Projected Net Ecosystem Productivity and Ecosystem Carbon Stocks over 250 years illustrate the comparative impacts of various disturbance, harvesting, and post-disturbance reforestation strategies in the Taiga Plains.** (a)Net Ecosystem Productivity (NEP, tC $ha^{-1}$) projected over a 250-year period under various disturbance and harvesting scenarios in the Taiga Plains. Scenarios include: BAU = Business-as-Usual; FFS = Future Fire Scenario; PH = Partial Harvesting; ClearCut = Clear Cut Harvesting; FFS_ClearCut_Planting = Future Fire Scenario with reforestation after clear cut harvesting; FFS_PH_Planting = Future Fire Scenario with reforestation after partial harvesting; BAU_PH_Planting = Business-as-Usual scenario with reforestation after partial harvesting; and BAU_ClearCut_Planting = Business-as-Usual scenario with reforestation after clear cut harvesting. Bars represent average NEP values across species and mixture types. The shaded regions around each line represent standard deviation, indicating variability in projected NEP across replicates. (b) Comparison of long-term Ecosystem Carbon Stocks (tC $ha^{-1}$) under different reforestation and disturbance recovery strategies in the Taiga Plains. Scenarios include: Business-as-Usual (BAU), Future Fire Scenario (FFS), and their respective variants with post-disturbance replanting following Partial Harvesting (PH) or Clear Cut harvesting. Results reflect simulated Ecosystem Carbon Stocks (tC $ha^{-1}$) over a 250-year simulation period.

## Net climate benefit considering albedo effects

Simulations integrating carbon sequestration and albedo-induced radiative forcing highlight critical trade-offs in assessing the net climate benefit of afforestation in the Taiga Plains ecoregion (Fig.5. (i), (ii)). Net $CO_2$-equivalent sequestration varied significantly across species mixtures and planting densities, with mixed-species configurations such as AllMix (25% SB/SW/Pine/AW) and 40% aspen (AW) achieving the highest net benefits under low-density conditions (~4.6 and 4.8 Mg $CO_2e$ $ha^{-1}$ $yr^{-1}$, respectively; Fig.5. (i)). These results persisted despite albedo-driven radiative offsets, which were quantified monthly in Table S2. For instance, 40% AW under low density exhibited pronounced cooling offsets in summer months (e.g., July: -0.466 Mg $CO_2e$ $ha^{-1}$ $yr^{-1}$), counteracting warming effects from reduced carbon sequestration efficiency.

In contrast, aspen monocultures (95% AW) demonstrated smaller annual albedo-driven cooling effects compared to mixed stands. Under low density, 95% AW showed moderate cooling offsets in March (-0.086 Mg $CO_2e$ $ha^{-1}$ $yr^{-1}$) but weaker effects in critical summer months like July (-0.466 Mg $CO_2e$ $ha^{-1}$ $yr^{-1}$; Table S2). However, its overall albedo offset profile (annual average: ~-0.21 Mg $CO_2e$ $ha^{-1}$ $yr^{-1}$) was less favorable than mixed stands, resulting in an ~11% reduction in climate mitigation efficacy due to residual warming penalties. Notably, coniferous monocultures (95% SB/SW) under medium density exhibited neutral to slightly positive albedo offsets (e.g., April: 0.310 Mg $CO_2e$ $ha^{-1}$ $yr^{-1}$ for 95% SW), amplifying their radiative trade-offs.

Monthly albedo offsets (March–November) were supplemented with December–February estimates from Hasler et al. (2023) [26] to approximate annual impacts (Table S2). Missing winter data (denoted by "--") reflect limitations in spatial interpolation, particularly for high-density stands. Deciduous-rich configurations (e.g., 40% AW) consistently delivered stronger cooling offsets (e.g., May: -0.321 Mg $CO_2e$ $ha^{-1}$ $yr^{-1}$ under low density), whereas mixed stands like AllMix balanced carbon gains with moderate albedo effects (e.g., April: 0.328 Mg $CO_2e$ $ha^{-1}$ $yr^{-1}$ under low density). These findings align with prior studies [20, 26], which emphasize biophysical feedbacks in boreal systems. However, our simulated albedo offsets were more conservative than those of Hasler et al. (2023) [26], likely due to site-specific snow duration and canopy structure differences (e.g., lower winter snowpack persistence reducing albedo contrasts).

Collectively, Table S2 underscores that optimizing species mixtures — prioritizing deciduous-rich or balanced mixes—and density regimes is essential to maximize net climate benefits when albedo effects are integrated. High-density coniferous stands, for example, showed minimal cooling (e.g., September: -0.02 Mg $CO_2e$ $ha^{-1}$ $yr^{-1}$ for 40% SB), reinforcing the need for strategic forest management in boreal regions.

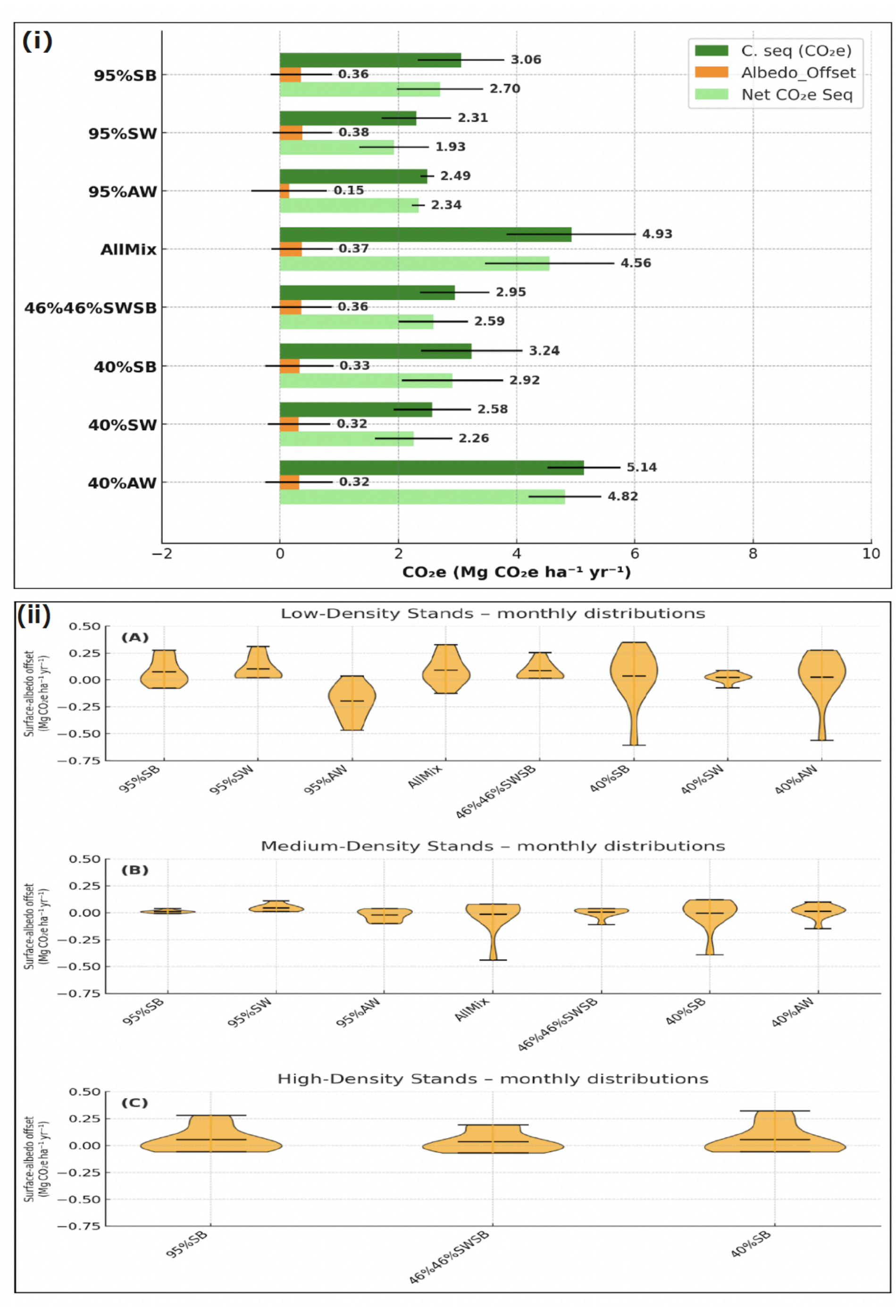

**Fig. 5. Integrated assessment of afforestation impacts in the Taiga Plains ecoregion, combining net climate benefits (carbon sequestration and albedo radiative forcing) with monthly albedo offset variability across species mixtures and planting densities.** (i) Net climate benefit of afforestation scenarios in the Taiga Plains ecoregion, expressed as Net $CO_2e$ Sequestration (Net $CO_2e$ Seq.), integrating carbon sequestration and albedo-induced radiative forcing effects. Values reflect a 100-year average (March–November), with December–January–February data supplemented from Hasler et al. (2023) [26]. Configurations include monocultures (95% Black spruce [SB], White spruce [SW], Aspen [AW]) and mixed stands (AllMix: 25% SB/SW/Pine/AW; 40% SB/SW/AW) under low, medium, and high planting densities. Simulations employ the CBM-CFS3 and GYPSY framework; error bands denote replicate variability. (ii) Monthly surface-albedo offsets ($Mg\ CO_2e\ ha^{-1}\ yr^{-1}$) for species-mixture scenarios across planting densities (A: Low, B: Medium, C: High). Violin plots depict distributions of nine monthly offsets per treatment (95% SB/SW/AW, AllMix, 46% SWSB, 40% SB/SW/AW), with orange fill, thick horizontal bars (mean), and thin whiskers (extrema). Positive values indicate net warming (reduced albedo); negative values denote cooling. Violin width corresponds to month-to-month variability.

## Validation of carbon stocks predictions

To verify model realism, we compared simulated above-ground biomass carbon (AGB C) trajectories against independent forest-inventory chronosequences [41, 42]. Overall, the validation confirms that the integrated framework captures boreal carbon accumulation patterns with high fidelity ($R^2 \approx 0.95–0.99$) (Fig. 11), lending confidence to our long-term projections. Collectively, these results demonstrate that moderate-density, mixed-species plantings yield the greatest climate benefit from boreal A/R. This strategy provides a scalable framework for climate-smart afforestation that can guide initiatives like Canada's 2 Billion Trees Program and inform broader climate mitigation efforts.

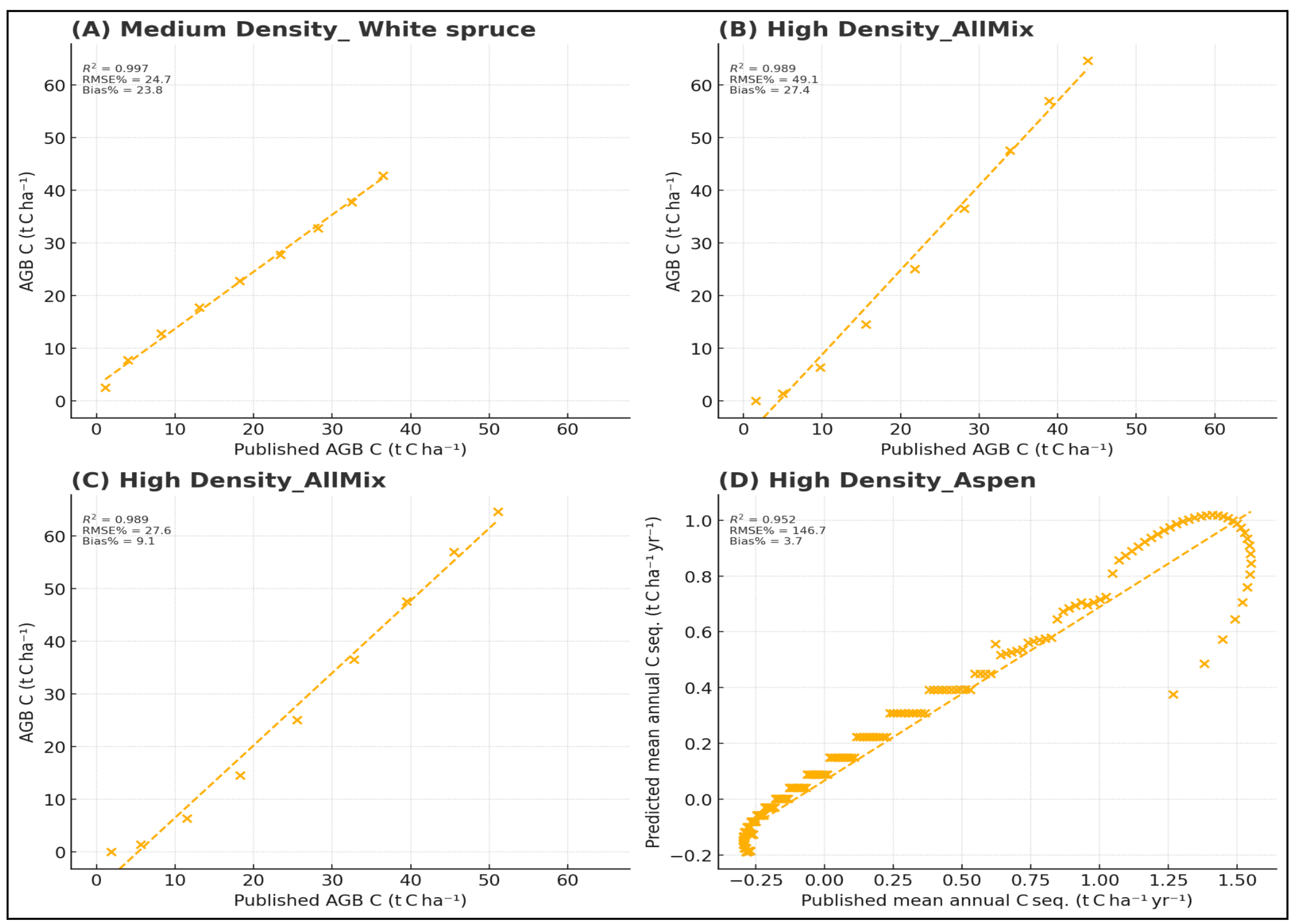

**Fig. 6. Validation of model predictions against independent chronosequences.** Scatterplots compare modelled values (y-axis) with published observations (x-axis) for four stand types in the Taiga Plains. Panels (A) – (C) depict above-ground biomass carbon (AGB C, t C $ha^{-1}$) for (A) medium-density white-spruce stands, (B) high-density AllMix vs. a 1 500 Asp + 1 000 Sw mixedwood, and (C) high-density AllMix vs. a 4 000 Asp + 1 000 Sw mixedwood; observations come from the Multisource Vegetation Inventory (MVI) satellite chronosequences [41]. Panel (D) validates mean annual carbon-sequestration rate (t C $ha^{-1}$ $yr^{-1}$) for high-density aspen against Landsat-derived yield-curves of [42]. Dashed lines are best-fit regressions; adjacent text gives coefficient of determination ($R^2$), root-mean-square error (RMSE %), and mean bias (Bias %). A common axis range is used in (A)–(C) to facilitate direct comparison. Slight over-prediction in (A) (+24 % bias) and in the mixedwood panels reflects the broader four-species composition of our AllMix treatment (25 % pine, 25 % white spruce, 25 % black spruce, 25 % aspen) relative to the two-species mixtures in the reference data.

# Discussion

Mixed-species stands consistently outperformed monocultures in long-term carbon accumulation and resilience. By year 250, stands blending deciduous and coniferous trees sequestered substantially more carbon than single-species stands and were better buffered against disturbances. Replanting under-stocked boreal forests with mixed species similarly enhanced carbon storage relative to letting those areas regenerate naturally. Crucially, biophysical feedbacks – notably surface albedo changes – reshaped the climate benefits of our scenarios: configurations dominated by deciduous trees maintained stronger net cooling effects than those dominated by dark, evergreen canopies. Model credibility is high: comparison with independent data showed that our simulations accurately reproduce observed biomass trajectories, increasing confidence in the robustness of these findings.

Notably**,** our 250-year dual carbon–albedo assessment goes beyond prior studies that examined shorter time frames or considered only carbon sequestration [27, 30]. By simulating ecosystem carbon dynamics *alongside* albedo-driven radiative forcing, we capture legacy effects and trade-offs that shorter or carbon-only analyses might miss. This integrated approach reinforces that maximizing carbon uptake alone is not sufficient – the species mix and stand structure must also be managed to minimize any offsetting warming from albedo change. Methodologically, the coupling of a forest growth model with an albedo forcing model is a key innovation of our study, allowing a direct comparison of biogeochemical and biophysical contributions to climate mitigation.

**Policy Implications for Global Boreal Afforestation**

Our findings have broad applicability to boreal afforestation efforts in other regions, such as Fennoscandia and Siberia. In high-latitude Eurasian forests, the net cooling benefit of tree planting similarly hinges on appropriate species selection and stand structure. For instance, afforesting open land with dense conifer plantations in snow-rich northern Europe can lead to net warming due to strong albedo reductions, whereas including deciduous species or using mixed stands yields a more favorable radiative outcome [30]. Large-scale model experiments indicate that realistic afforestation in boreal Asia would likely produce only modest cooling [38,39], reinforcing the importance of maximizing carbon gains to overcome the albedo penalty. In the context of the IPCC's global mitigation frameworks, these insights demonstrate that the climate return on tree-planting investments will be greatest when both carbon sequestration and albedo are considered [40]. Our results provide a concrete strategy – favoring mixed-species, intermediate-density plantings – that can be incorporated into national reforestation programs. By adopting such climate-smart practices, policymakers in Canada and across the boreal zone (e.g., Scandinavia and Russia) can ensure that ambitious tree-planting campaigns translate into lasting net cooling contributions.

**Limitations and Future Direction**

This study is focused on one ecoregion (the Taiga Plains) and assumes fixed climatic conditions; other boreal regions with different climates or site conditions may exhibit different carbon–albedo dynamics. Likewise, our models do not account for future climate changes that could affect tree growth and snow cover, nor for certain feedbacks like permafrost thaw or changing disturbance regimes beyond the scenarios considered. We also did not incorporate socio-economic factors – such as land-use constraints, costs, or management policies – which will influence the real-world feasibility of implementing our recommendations. Future work should integrate climate projections, economic analyses, and Indigenous and local stakeholder perspectives to identify where mixed-species strategies are most viable and beneficial. Cross-regional validation of our integrated carbon–albedo approach (for example, through pilot projects in both North American and Eurasian boreal forests) would strengthen the generality of our conclusions. Despite these limitations, our dual modeling framework offers a valuable tool for designing boreal afforestation initiatives that achieve enduring climate benefits while minimizing unintended warming effects.

# Methods

## Study Area

The Taiga Plains ecoregion spans approximately 570,000 km[2] across southwestern Northwest Territories and parts of northeastern British Columbia and Alberta. The area features discontinuous permafrost, low-relief topography, and is dominated by black spruce, white spruce, and trembling aspen [16, 17]. Wetlands such as fens and bogs occupy roughly 25% of the landscape and are crucial for long-term carbon storage [16]. Climate change has increased fire severity and reduced intervals between fire events, thereby accelerating shifts in vegetation composition [10, 21].

## Scenario Development

We designed a factorial experiment using three planting densities: low (≤ 600 trees/ha), medium (600 – 1400 trees/ha), and high (1400 – 2500 trees/ha). Eight species configurations were modeled, including three monocultures (95.2% dominance) of black spruce, white spruce, and aspen, and five mixed-species scenarios (e.g., AllMix: 25% of each species) [10, 15, 43] (Table 1). These scenarios reflect plausible reforestation strategies observed in post-disturbance boreal ecosystems [15].

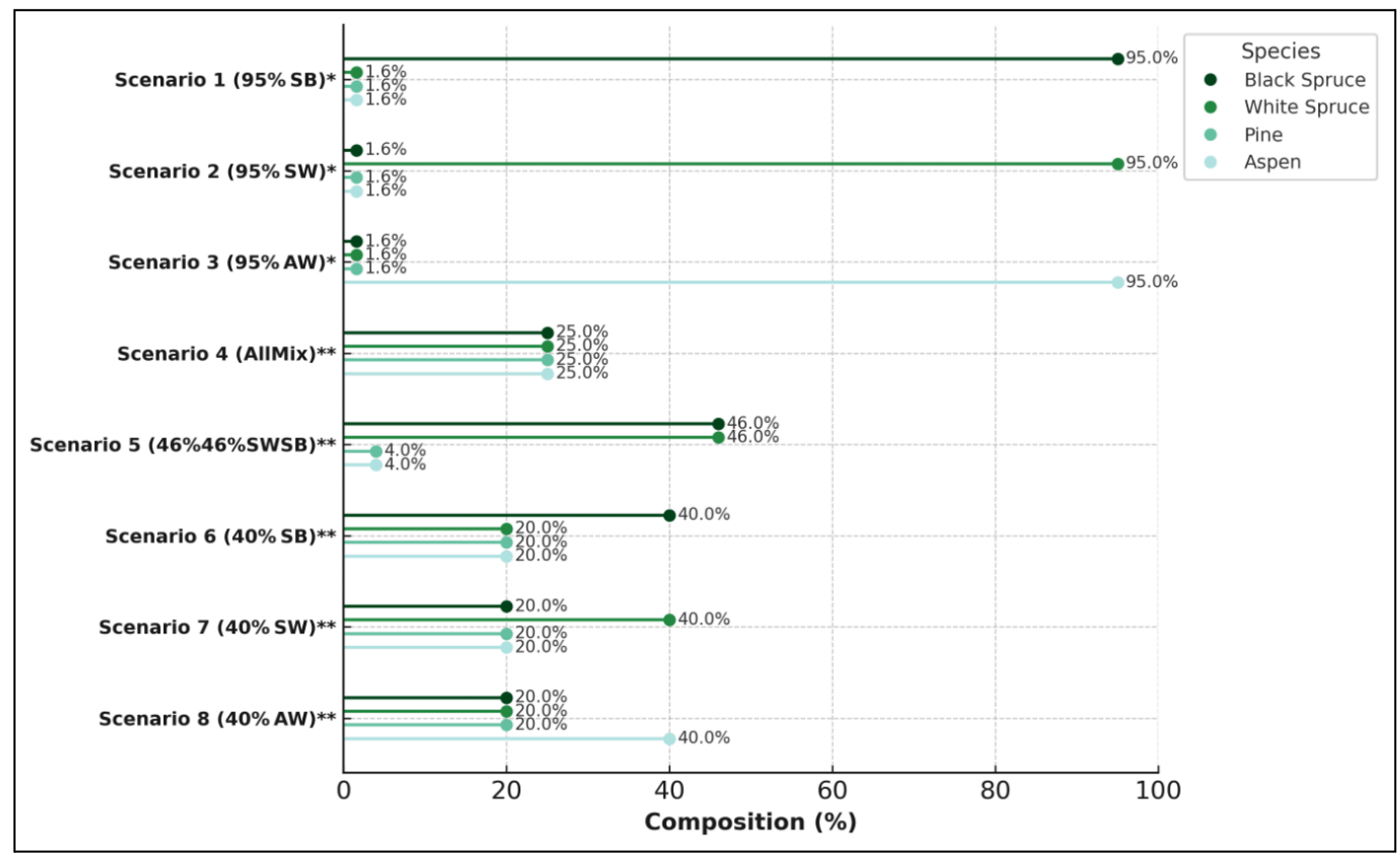

**Fig. 7:** Forest stand species composition scenarios. * Monoculture ** Mixed-species

## Forest Growth Modelling

GYPSY [23] was employed to simulate growth trajectories for white spruce, black spruce, and aspen, calibrated with site-specific data from Taiga Plains studies [8, 10, 44]. Yield projections incorporated merchantable volume formulas based on DBH, height, and species-specific taper functions:

$$\mathbf{V = a + b \cdot D^{c} \cdot H^{d}} \qquad (1)$$

Where **V** is tree volume, **D** is diameter at breast height (DBH), **H** is height, and **a**, **b**, **c**, and **d** are empirically derived constants. GYPSY simulated biomass accumulation across all combinations of species composition and density.

Forest stand structural dynamics and various configurations were additionally informed by empirical insights and operational benchmarks outlined in regional forest management plans such as [10, 43]. This helped ensure the simulations reflect realistic forest development pathways observed under management conditions in boreal environments.

### Carbon Simulation with CBM-CFS3

The CBM-CFS3 model [24] calculated carbon stocks and fluxes in five ecosystem pools: aboveground biomass, belowground biomass, litter, deadwood, and soil organic carbon (SOC). Total ecosystem carbon (TEC) was estimated as:

$$\mathbf{TEC}(t) = \mathbf{C_{AGB}}(t) + \mathbf{C_{BGB}}(t) + \mathbf{C_{DOM}}(t) + \mathbf{C_{LIT}}(t) + \mathbf{C_{SOC}}(t) \qquad (2)$$

Where:
$\mathbf{C_{AGB}}$ = Aboveground biomass carbon
$\mathbf{C_{BGB}}$ = Belowground biomass carbon
$\mathbf{C_{DOM}}$ = Dead organic matter (e.g., standing deadwood)
$\mathbf{C_{LIT}}$ = Litter
$\mathbf{C_{SOC}}$ = Soil organic carbon
$t$ = simulation year

Stand-level simulations were initialized using the Stand-Level Project Creator (SLPC), while large-scale landscape dynamics were captured with the Standard Import Tool (SIT). Disturbances modeled included fire (based on [45, 46]), insect outbreaks, and two harvesting treatments: clearcut (97% removal) and partial harvest (50%) [24, 47].

These data informed fire return intervals, spatial burn distribution, and species-specific susceptibility, allowing simulations to reflect both baseline and climate-enhanced fire regimes and improve the realism of carbon flux estimates across afforestation and reforestation scenarios.

### Carbon Flux Metrics and Equations

The CBM-CFS3 model computed the following key carbon flux metrics throughout the 250-year simulations:

Net Primary Productivity (NPP) – the net carbon assimilated by vegetation after autotrophic respiration:

$$\mathbf{NPP}(t) = \mathbf{GPP}(t) - \mathbf{R_a(t)} \quad (3)$$

Where, **GPP(t)** = gross primary productivity (total carbon fixed through photosynthesis), $\mathbf{R_a(t)}$ = autotrophic respiration (carbon expended for plant metabolism). In CBM-CFS3, NPP is determined using species growth curves and modified by climatic variables (e.g., warming, permafrost thaw).

**NEP(t)** = Net Ecosystem Productivity at time *t* (i.e., net annual carbon sequestration rate, in t C $ha^{-1}$ $yr^{-1}$). It is the overall carbon balance of the ecosystem, calculated as NPP minus heterotrophic respiration (Rh) from microbial decomposition and other carbon losses such as disturban emission:

$$\mathbf{NEP}(t) = \mathbf{NPP}(t) - \mathbf{R_h}(t) \quad (4)$$

Positive NEP values indicate the ecosystem acts as a carbon sink, while negative values denote a carbon source.

Model setup and parameterization followed CBM-CFS3 guidelines [24], with site-specific adjustments to account for boreal climate dynamics, such as disturbance dynamics, soil characteristics and plant functional traits in the Taiga Plains.

**Albedo effects estimation**

Daily surface albedo data were obtained from [29] and transformed to mean monthly values on a spatial grid of 0.05-degree resolution. Monthly all-sky shortwave radiative kernel data, representing the change in top-of-atmosphere (TOA) radiative flux per unit change in surface albedo (W/m[2]/albedo), were acquired from [30]. To ensure spatial consistency, kernel data were bilinearly interpolated to the albedo dataset resolution. Forest inventory data (comprising grid cell coordinates, forest percentage, dominant species composition, and forest age) were obtained from the Canadian National Terrestrial Ecosystem Monitoring System – Stand-Based Forest Inventory (NTEMS-SBFI). Analysis was restricted to grid cells within the Taiga Plains ecoregion.

**Forest Scenario Definition and Assignment**

Each grid cell was categorized under low (< 24%), medium (24 – 56%), or high (>56%) forest cover, corresponding with Low-density stands (≤ 600 trees/ha), medium densities (600–1400 stems/ha) and High-density stands (1400–2500 trees/ha), respectively. Eight forest composition scenarios were defined using relative proportions of four major species groups: Black Spruce, White Spruce, Pine, and Aspen (Table 1). Scenario assignment was based on species dominance or compositional similarity using the least absolute difference approach. Species in the inventory data were mapped into the four reference categories based on ecological and structural traits:

1. Broadleaf deciduous trees and deciduous conifers (e.g., larches) were assigned to Aspen.
2. Pines were grouped under Pine.
3. Spruce species were split into Black spruce or White spruce categories based on regional distributions.
4. Douglas-fir was mapped to the White Spruce category.

This approach ensured consistency in assigning forest types to each grid cell. Cells with invalid forest percentage data were excluded.

**Albedo Parameter Estimation**

Monthly albedo and kernel values were extracted and spatially averaged for each grid cell, generating 12-month time series. A linear mixture model was used to estimate intrinsic albedo values:

$$\text{Albedo_observed}(t) = \text{P_forest} * \text{A_forest}(s, t) + (1 - \text{P_forest}) * \text{A_free}(t) \quad (5)$$

Where: $t$ = month (1–12), $s$ = scenario index (0–7), A_forest(s, t) = albedo of scenario s, A_free(t) = albedo of non-forested land.

Parameters were estimated using the Trust Region Reflective algorithm via scipy.optimize.least_squares, with bounds set to [0, 1].

**Radiative Forcing and $CO_2$ Equivalent Calculation**

The monthly change in albedo (ΔAlbedo(s, t)) was used to compute the TOA radiative forcing (RF) per scenario:

$$RF(s,t) = \Delta\text{Albedo}(s,t) \cdot \text{Kernel mean}\ (s,t) \quad (6)$$

Monthly albedo data [29] and radiative kernels [30] were combined to estimate radiative forcing (RF) for each scenario. RF was converted to $CO_2e$ using methodologies from [31], with winter gaps filled using their boreal-specific estimates. RF values ($W/m^2$) were then converted to Mg $CO_2e$ $ha^{-1}$ $yr^{-1}$ following [31] using the following steps:

1. Scale RF regionally over 25,200 ha.
2. Convert RF to global atmospheric carbon change using: ΔC_CO2e = (2.13 Pg C/ppm * 400 ppm) * [exp(RF_global / 5.35) − 1]
3. Convert Pg C to Pg $CO_2$ using a factor of 3.67.
4. Convert Pg $CO_2$ to Mg $CO_2$.
5. Normalize by regional area.
6. Adjust using a 100-year horizon factor (f_100 = 0.40).

## Validation and Statistics

Model predictions were validated against satellite-derived chronosequences and field-based biomass estimates from comparable boreal forests in the Taiga Plains [41, 42]. Simulated aboveground biomass carbon (AGB C) for representative species–density configurations matched observed data with high fidelity ($R^2$ = 0.95–0.99), and root mean square error (RMSE) ranged between 5–11%, indicating excellent predictive performance. Bias values were generally within ±25%, with slight over-predictions in some mixed-wood comparisons due to differences in species compositions between modeled and observed stands (e.g., AllMix vs. 2-species inventories) [41].

To assess treatment effects statistically, we conducted two-way ANOVA tests on Net Ecosystem Carbon (NEC) at years 50, 100, and 250, examining interactions between species composition and planting density. Post-hoc analysis using Tukey's Honest Significant Difference (HSD) test ($\alpha = 0.05$) identified significant differences among treatment groups. Visual matrices (Fig. 2iii, Fig. 3ii) highlighted significant pairwise contrasts using 1 − p-value heatmaps, with red asterisks denoting statistical significance ($p < 0.05$). These results confirm that moderate-density, mixed-species configurations generally outperform monocultures and BAU scenarios in achieving higher NEC by year 250.

**Data Availability**
The data that support the findings of this study are publicly available to download and are referenced in the bibliography. Refer to the Methods section for more details.

**Acknowledgements**
We thank the Natural Sciences and Engineering Research Council of Canada (NSERC) for primarily funding this research through the NSERC Alliance Mission grant - ALLRP 577126-2022 (Y.L., R.B., and J.M.-C.).

**Author Contributions**
E.O. conducted the research, literature review, performed simulations, wrote the article, plotted results, and created illustrations. K.B.D. helped with ideation, literature review, data, visualization and editing. R.B., P. M. ,J.M.-C., and Y.L. were involved in the acquisition of funding, editing of the manuscript and overall supervision. All authors contributed to discussion and conceptualization of arguments.

**Competing Interests**
The authors declare no competing interests.

**Supplementary Information**
Attached as a separate supplementary information.

# Supplementary

## Growth and Yield Curves

Analysis of merchantable volume trajectories provided critical insights for operational silviculture and long-term forest carbon planning in boreal systems. Results from the Growth and Yield Projection System (GYPSY) simulations (Fig. S1 & S2) indicated that aspen (*Populus tremuloides*) monocultures exhibited the highest early-phase merchantable volume yields, peaking at approximately 300 $m^3$ $ha^{-1}$ by Year 50. This trend closely paralleled their rapid Net Ecosystem Carbon (NEC) accumulation during the first century (Fig. S1 & S2), consistent with aspen's known physiology as a shade-intolerant, fast-growing pioneer species that capitalizes on early light availability and nutrient pulses following disturbance [13, 14].

However, volume yields in aspen-dominated stands declined markedly in the second century of simulation, with a near plateau in both NEC and biomass carbon by Year 150. This decline is attributable to aspen's relatively short life span, susceptibility to decay and biotic stressors, and limited capacity for self-regeneration under closed canopy conditions, factors frequently observed in post-disturbance boreal chronosequences [13, 14]. In contrast, mixed-species stands, particularly those with balanced proportions of black spruce (*Picea mariana*), white spruce (*Picea glauca*), pine (genus, *pinus*), and aspen, demonstrated more stable and resilient yield profiles over time. At medium planting densities (600–1400 stems $ha^{-1}$), these configurations sustained merchantable volume gains well beyond Year 150, while simultaneously achieving higher cumulative NEC than any monoculture by Year 250 (Fig. S1 & S2). This outcome reflects interspecific complementarity, whereby rapid early growth by aspen is compensated by the shade-tolerant, stress-resilient biomass retention of spruce and pine in later years [40, 41].

These findings align with earlier modeling and empirical work indicating that moderate-density, mixed-species stands can optimize both timber supply and ecosystem service delivery across multi-century horizons [23]. From a forest management perspective, this suggests that species diversification does not inherently compromise timber productivity. On the contrary, it may buffer long-term yield losses associated with disturbance, senescence, or climate variability, thus supporting both economic and ecological resilience in boreal afforestation systems [9].

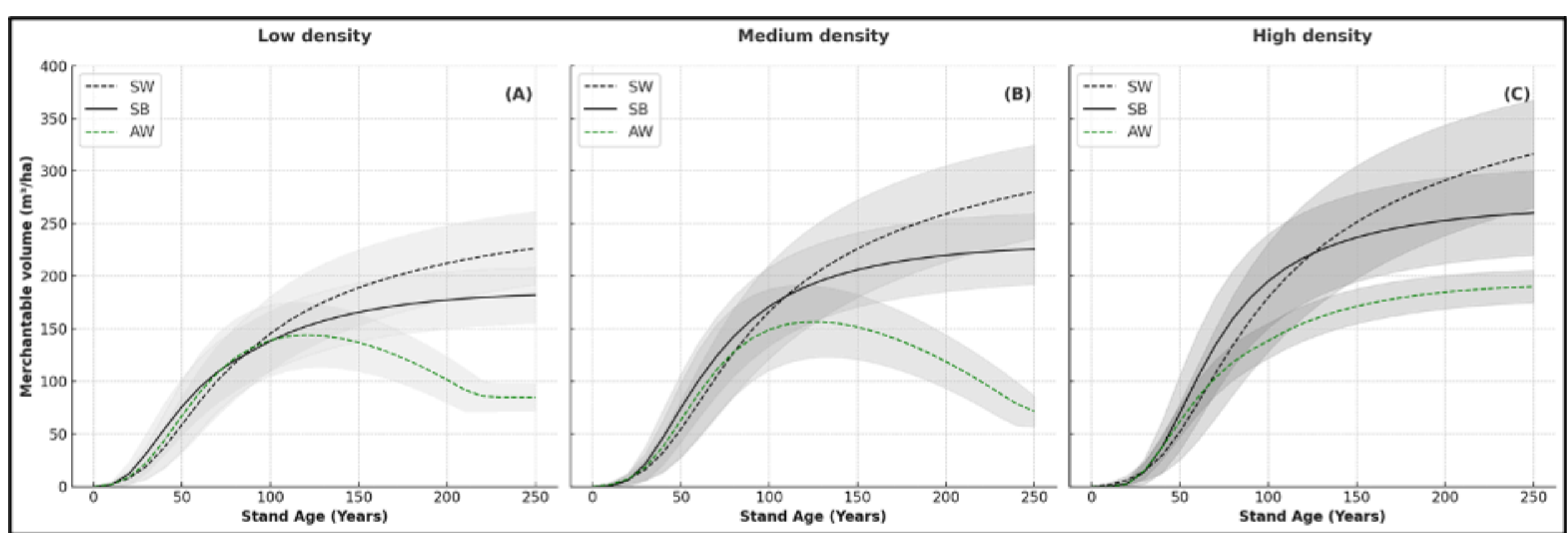

**Fig. S1.** Merchantable volume ($m^3$/ha) of White spruce (SW), Black spruce (SB), and Aspen (AW) in low (A), medium (B) and high (C) density stands over a 250-year period in the Taiga Plains within the Northwestern Territory of Canada. The curves were developed using the Growth and Yield Projection System (GYPSY). The shaded regions represent the spread of standard deviation (SD) for each species, indicating variability in projected growth trends.

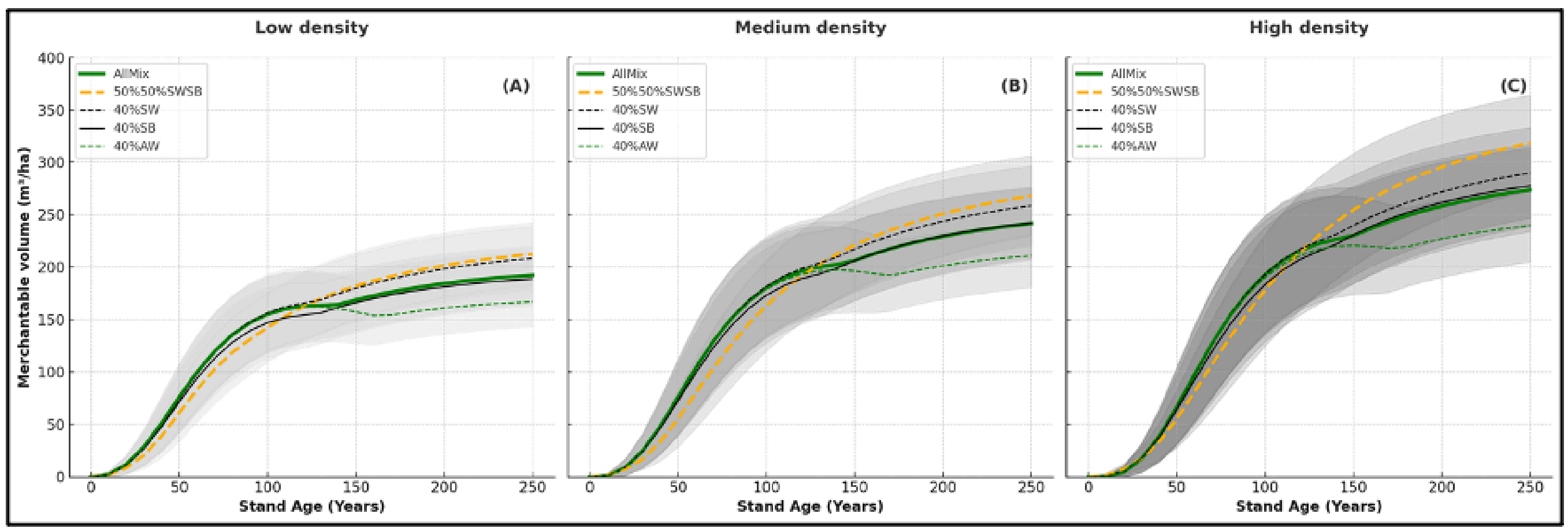

**Fig. S2.** Merchantable volume (m³/ha) projections for mixed-species stands across low (A), medium (B), and high (C) planting densities over a 250-year simulation period in the Taiga Plains ecoregion. Outputs were generated using the Growth and Yield Projection System (GYPSY) with species-specific inputs for white spruce, black spruce, aspen, and pine. Shaded regions represent standard deviation across replicates, illustrating variability in stand-level yield outcomes under different species combinations and density regime.

## Annual carbon Sequestration potential of various species configuration

Annual carbon sequestration potential in different species mixtures across low, medium, and high stand densities (Fig. S3*)* shows that carbon sequestration peaks between Years 40–80 across all scenarios. Beyond Year 200, the curves plateau, indicating saturation of biomass and stabilization of net carbon fluxes.

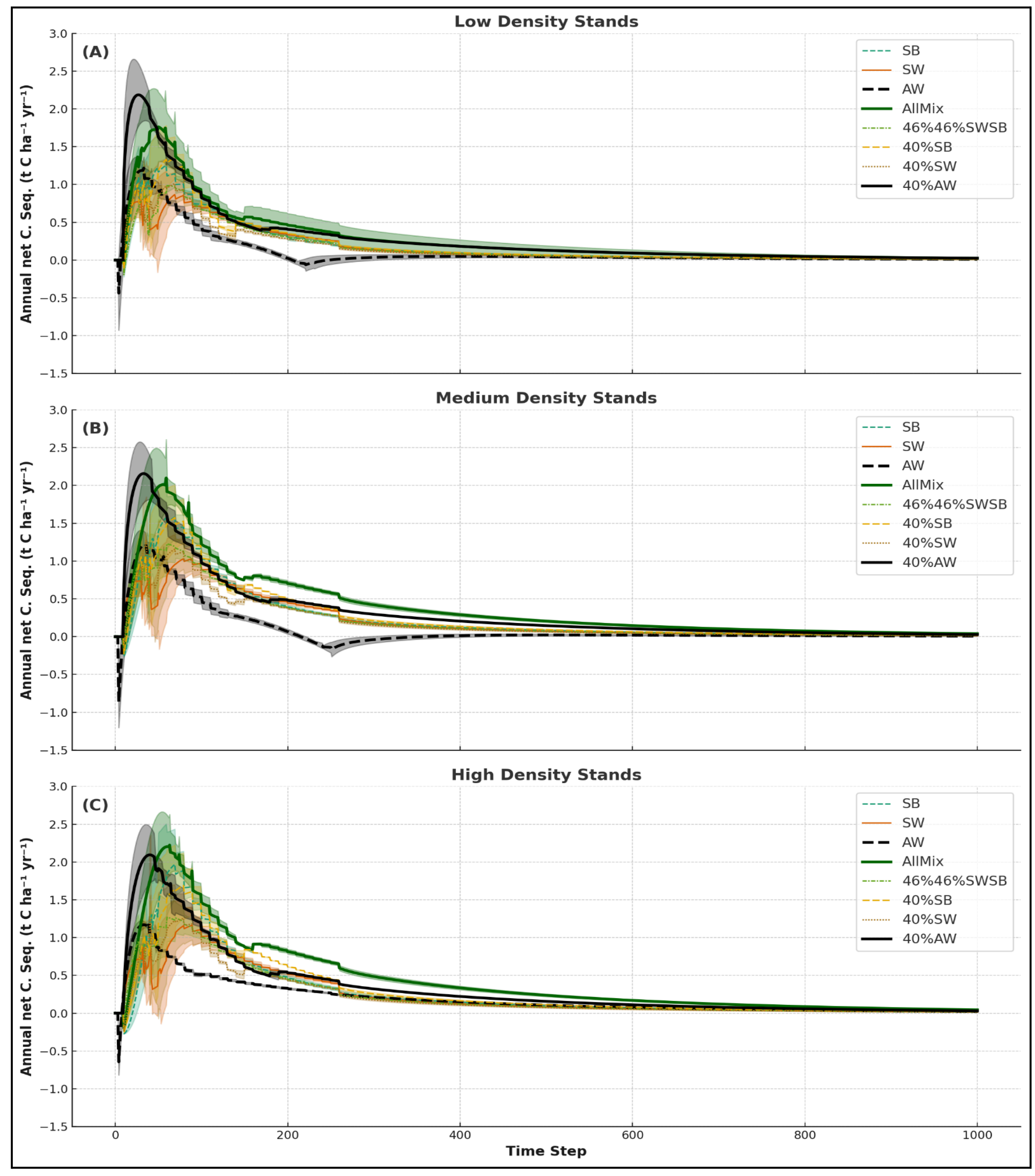

**Fig. S3.** Annual net carbon sequestration trajectories under varying species compositions and planting densities**.**Annual net carbon sequestration ($t\ C\ ha^{-1}\ yr^{-1}$) is shown over a 1000-year simulation period in the Taiga Plains ecoregion for eight afforestation scenarios representing different species compositions (including monocultures and mixed-species stands) under three planting density regimes. Panels A, B, and C correspond to low, medium, and high planting densities, respectively, with each panel displaying the sequestration trajectories of all eight species composition scenarios. Carbon fluxes were simulated using the Carbon Budget Model (CBM-CFS3), driven by growth projections from the GYPSY model. Each colored line represents the mean annual sequestration rate for a given stand composition at the specified planting density, and the shaded band around each line indicates ±1 standard deviation across simulation replicates, illustrating the variability in long-term carbon uptake for that scenario.

## Biomass carbon trends of various species configuration

Cumulative biomass carbon stocks over 1000 years for each planting density category (Fig. S4) underscores that mixed-species stands under medium and high densities maintain consistently higher biomass stocks, with a visible tapering off of accumulation after Year 200.

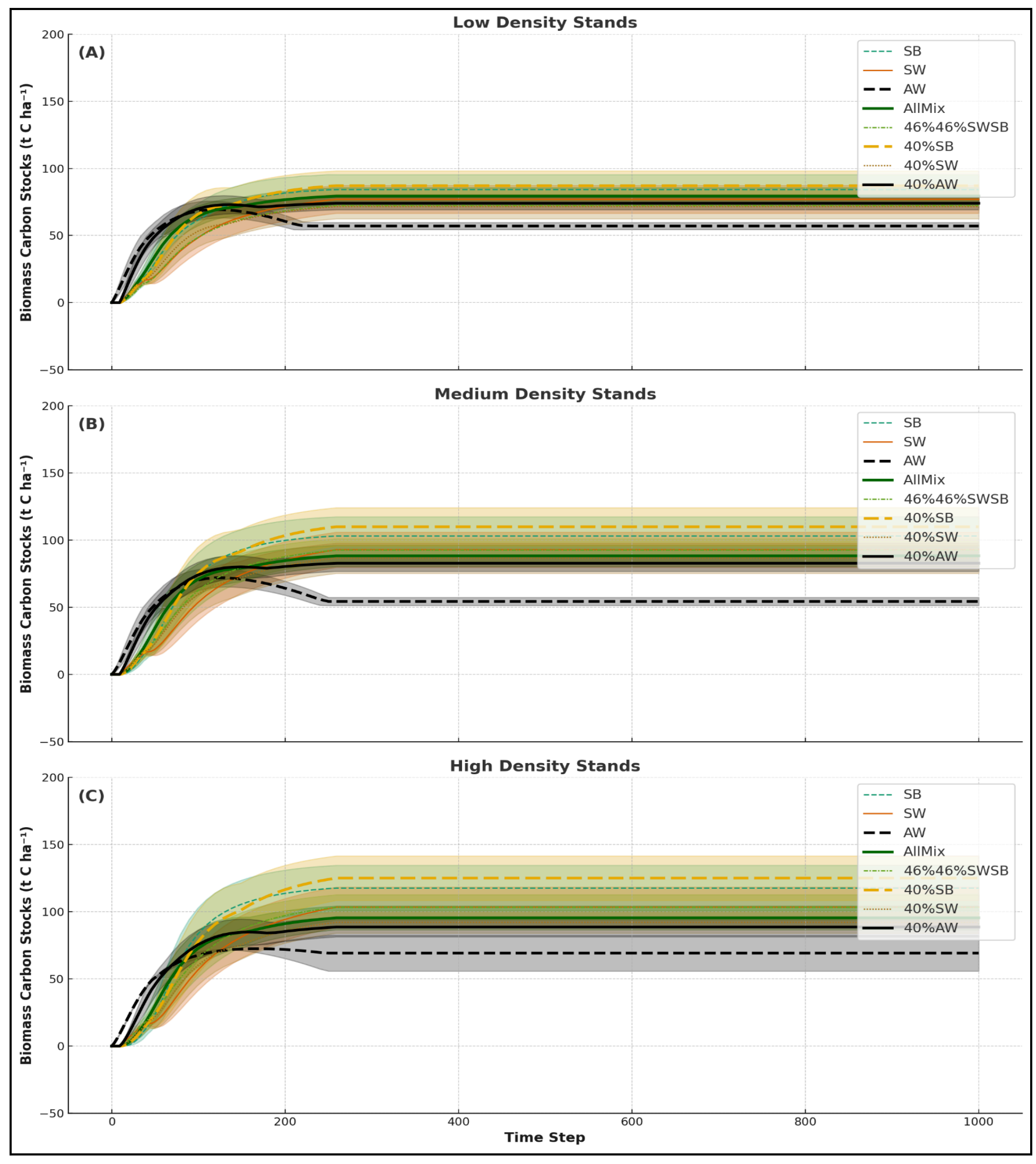

**Fig. S4.** Biomass carbon stock trajectories over a 1000-year simulation period across varying species compositions and planting densities. The figure displays above- and belowground biomass carbon stocks (t C $ha^{-1}$) for eight forest stand scenarios—including monocultures and mixed-species configurations—under three planting density regimes: low (A), medium (B), and high (C). Simulations were conducted using the GYPSY growth model and CBM-CFS3 carbon accounting framework. Each line represents the mean biomass carbon stock for a given stand composition and density, while the shaded areas around the lines denote ±(standard deviation), capturing variability across replicates.

**Table S1**. Average annual carbon sequestration rates (t C $ha^{-1}$ $yr^{-1}$) and corresponding $CO_2$ equivalent ($CO_2e$) values (t $CO_2e$ $ha^{-1}$ $yr^{-1}$) over a 100-year period for eight afforestation scenarios. Each scenario (table row) represents a distinct species composition – ranging from monocultures (e.g., 95% black spruce (SB)) to mixed stands (AllMix), evaluated under three planting density levels (columns: low, medium, high). $CO_2$ equivalent values were derived using a standard conversion factor of 1 t C = 3.67 t $CO_2e$.

| Scenarios/Treatments | Low Density Stands (≤ 600 trees/ha.) | | Medium Density Stands (600 - 1400 trees/ha.) | | High Density Stands (1400 - 2500 trees/ha.) | |
|---|---|---|---|---|---|---|
| | Carbon seq. rate (t C $ha^{-1}$ $yr^{-1}$) | $CO_2e$ Seq (t $CO_2e$ $ha^{-1}$ $yr^{-1}$) | Carbon seq. rate (t C $ha^{-1}$ $yr^{-1}$) | $CO_2e$ Seq (t $CO_2e$ $ha^{-1}$ $yr^{-1}$) | Carbon seq. rate (t C $ha^{-1}$ $yr^{-1}$) | $CO_2e$ Seq (t $CO_2e$ $ha^{-1}$ $yr^{-1}$) |
| SB | 0.83(±0.20) | 3.1(±0.73) | 0.98(±0.25) | 3.61(±0.92) | 1.90(±0.29) | 3.98(±1.08) |
| SW | 0.63(±0.16) | 2.3(±0.59) | 0.69(±0.19) | 2.52(±0.71) | 0.73(±0.22) | 2.66(±0.80) |
| AW | 0.69(±0.03) | 2.5(±0.11) | 0.71(±0.08) | 2.59(±0.30) | 0.71(±0.07) | 2.60(±0.27) |
| AllMix | 1.3(±0.23) | 4.9(±1.1) | 1.36(±0.21) | 5.01(±0.77) | 1.42(±0.30) | 5.22(±1.08) |
| 46%46%SWSB | 0.81(±0.16) | 3.0 (±0.59) | 0.85(±0.19) | 3.10(±0.68) | 0.88(±0.21) | 3.23(±0.78) |
| 40%SB | 0.90(±0.23) | 3.2(±0.86) | 0.99(±0.30) | 3.65(±0.78) | 1.01(±0.34) | 3.70(±1.26) |
| 40%SW | 0.70(±0.18) | 2.6(±0.66) | 0.78(±0.21) | 2.86(±0.68) | 0.79(±0.25) | 2.91(±0.90) |
| 40%AW | 1.4(±0.17) | 5.1(±0.62) | 1.5(±0.20) | 5.32(±0.72) | 1.45(±0.22) | 5.31(±0.81) |

**Table S2.** Monthly albedo offset values expressed as $CO_2$-equivalent (Mg $CO_2e$ $ha^{-1}$ $yr^{-1}$) for eight afforestation scenarios under low, medium, and high planting densities. Each scenario corresponds to a distinct forest composition (monoculture and mixed-species configurations). The values represent the estimated radiative forcing due to albedo change as outlined in the 'Albedo effects estimation' section of the Methods. Dashes (–) indicate missing data due to limitations in the input datasets or spatial inconsistency during interpolation and scenario assignment.

| | TREATMENT/SCENARIOS | | | | | | | |
|---|---|---|---|---|---|---|---|---|
| | **95%SB** | **95%SW** | **95%AW** | **AllMix** | **46%46%SWSB** | **40%SB** | **40%SW** | **40%AW** |
| **LOW DENSITY STANDS** | | | | | | | | |
| Albedo offset_January | - | - | - | - | - | - | - | - |
| Albedo offset_February | - | - | - | - | - | - | - | - |
| Albedo offset_March | 0.147 | 0.167 | -0.086 | 0.170 | 0.145 | 0.191 | 0.045 | 0.135 |
| Albedo offset_April | 0.276 | 0.310 | -0.144 | 0.328 | 0.255 | 0.348 | 0.058 | 0.276 |
| Albedo offset_May | 0.253 | 0.231 | -0.321 | 0.251 | 0.153 | 0.236 | -0.073 | 0.232 |
| Albedo offset_June | 0.065 | 0.088 | -0.344 | 0.088 | 0.088 | 0.086 | 0.088 | 0.085 |
| Albedo offset_July | -0.006 | 0.052 | -0.466 | 0.041 | 0.031 | 0.022 | 0.028 | 0.026 |
| Albedo offset_August | -0.004 | 0.023 | -0.178 | 0.021 | 0.015 | 0.011 | 0.013 | 0.012 |
| Albedo offset_September | 0.000 | 0.021 | -0.181 | 0.019 | 0.018 | 0.011 | 0.013 | 0.012 |
| Albedo offset_October | 0.009 | 0.020 | -0.098 | 0.022 | 0.019 | 0.023 | 0.018 | 0.022 |
| Albedo offset_November | -0.078 | 0.031 | 0.037 | -0.126 | 0.037 | -0.607 | 0.023 | -0.562 |
| Albedo offset_December | - | - | - | - | - | - | - | - |
| **MEDIUM DENSITY STANDS** | | | | | | | | |
| Albedo offset_January | - | - | - | - | - | - | - | - |
| Albedo offset_February | - | - | - | - | - | - | - | - |
| Albedo offset_March | 0.02 | 0.07 | -0.03 | 0.02 | 0.00 | 0.04 | - | 0.03 |
| Albedo offset_April | 0.04 | 0.11 | -0.07 | 0.07 | -0.01 | 0.10 | - | 0.05 |
| Albedo offset_May | 0.00 | 0.05 | -0.09 | 0.08 | -0.11 | 0.12 | - | 0.10 |
| Albedo offset_June | 0.01 | 0.06 | -0.10 | 0.02 | 0.04 | 0.03 | - | 0.03 |
| Albedo offset_July | 0.01 | 0.03 | 0.04 | 0.05 | 0.03 | 0.02 | - | 0.02 |
| Albedo offset_August | 0.01 | 0.01 | 0.02 | 0.03 | 0.02 | 0.01 | - | 0.01 |
| Albedo offset_September | 0.01 | 0.02 | 0.02 | 0.02 | 0.02 | 0.01 | - | 0.01 |
| Albedo offset_October | 0.00 | 0.01 | 0.01 | 0.02 | 0.01 | 0.01 | - | 0.01 |
| Albedo offset_November | -0.01 | 0.04 | -0.01 | -0.44 | 0.04 | -0.39 | - | -0.15 |
| Albedo offset_December | - | - | - | - | - | - | - | - |
| **HIGH DENSITY STANDS** | | | | | | | | |
| Albedo offset_January | - | - | - | - | - | - | - | - |
| Albedo offset_February | - | - | - | - | - | - | - | - |
| Albedo offset_March | 0.05 | - | - | - | 0.03 | 0.22 | - | - |
| Albedo offset_April | 0.20 | - | - | - | 0.18 | 0.32 | - | - |
| Albedo offset_May | 0.28 | - | - | - | 0.19 | 0.08 | - | - |
| Albedo offset_June | 0.06 | - | - | - | 0.06 | -0.06 | - | - |
| Albedo offset_July | -0.06 | - | - | - | -0.07 | -0.04 | - | - |
| Albedo offset_August | -0.03 | - | - | - | -0.04 | -0.02 | - | - |
| Albedo offset_September | -0.02 | - | - | - | -0.02 | -0.01 | - | - |
| Albedo offset_October | -0.01 | - | - | - | -0.02 | -0.01 | - | - |
| Albedo offset_November | 0.00 | - | - | - | 0.00 | -0.01 | - | - |
| Albedo offset_December | - | - | - | - | - | - | - | - |